\documentclass[aps,prl,twocolumn,letterpaper,superscriptaddress,showpacs]{revtex4}
\usepackage{graphicx}

\begin{document}

\title{
Itinerant Antiferromagnetism in RuO$_{2}$
}
\author{T. Berlijn}
\affiliation{Center for Nanophase Materials Sciences, Oak Ridge National Laboratory, Oak Ridge, Tennessee 37831}
\affiliation{Computer Science and Mathematics Division, Oak Ridge National Laboratory, Oak Ridge, Tennessee 37831}
\author{P. C. Snijders}
\affiliation{Materials Science and Technology Division, Oak Ridge National Laboratory, Oak Ridge, Tennessee 37831}
\affiliation{Department of Physics and Astronomy, The University of Tennessee, Knoxville, Tennessee 37996}
\author{O. Delaire}
\affiliation{Materials Science and Technology Division, Oak Ridge National Laboratory, Oak Ridge, Tennessee 37831}
\affiliation{Mechanical Engineering and Materials Science, Duke University, Durham, North Carolina 27708, USA}
\author{H.-D. Zhou}
\affiliation{Department of Physics and Astronomy, The University of Tennessee, Knoxville, Tennessee 37996}
\author{T. A. Maier}
\affiliation{Center for Nanophase Materials Sciences, Oak Ridge National Laboratory, Oak Ridge, Tennessee 37831}
\affiliation{Computer Science and Mathematics Division, Oak Ridge National Laboratory, Oak Ridge, Tennessee 37831}
\author{H.-B. Cao}
\affiliation{Quantum Condensed Matter Division, Oak Ridge National Laboratory, Oak Ridge, Tennessee 37831}
\author{S.-X. Chi}
\affiliation{Quantum Condensed Matter Division, Oak Ridge National Laboratory, Oak Ridge, Tennessee 37831}
\author{M. Matsuda}
\affiliation{Quantum Condensed Matter Division, Oak Ridge National Laboratory, Oak Ridge, Tennessee 37831}
\author{Y. Wang}
\affiliation{Materials Science and Technology Division, Oak Ridge National Laboratory, Oak Ridge, Tennessee 37831}
\author{M. R. Koehler}
\affiliation{Department of Materials Science and Engineering, The University of Tennessee, Knoxville, Tennessee 37996}
\author{P. R. C. Kent}
\affiliation{Center for Nanophase Materials Sciences, Oak Ridge National Laboratory, Oak Ridge, Tennessee 37831}
\affiliation{Computer Science and Mathematics Division, Oak Ridge National Laboratory, Oak Ridge, Tennessee 37831}
\author{H. H. Weitering}
\affiliation{Department of Physics and Astronomy, The University of Tennessee, Knoxville, Tennessee 37996}
\affiliation{Materials Science and Technology Division, Oak Ridge National Laboratory, Oak Ridge, Tennessee 37831}

\date{\today}

\begin{abstract}
Bulk rutile RuO$_2$ has long been considered a Pauli paramagnet. Here we report that RuO$_2$ exhibits a hitherto undetected lattice distortion below approximately 900 K. The distortion is accompanied by antiferromagnetic order up to at least 300 K with a small room temperature magnetic moment of approximately 0.05 $\mu_B$ as evidenced by polarized neutron diffraction. Density functional theory plus $U$ (DFT+$U$) calculations indicate that antiferromagnetism is favored even for small values of the Hubbard $U$ of the order of 1 eV. The antiferromagnetism may be traced to a Fermi surface instability, lifting the band degeneracy imposed by the rutile crystal field. The combination of high N\'eel temperature and small itinerant moments make RuO$_2$ unique among ruthenate compounds and among oxide materials in general.

\end{abstract}


\maketitle

Theories of magnetism in $3d$ transition metal oxides (TMOs) are usually framed in the context of strong Coulomb repulsions and Hund's rule coupling in the $3d$ orbitals of the transition metal cation, and their covalent bonding with the oxygen $2p$ orbitals.
Strong on-site electron interactions tend to inhibit double occupancy of the $3d$ orbital and the overall Coulomb energy of the crystal is lowered by localizing the valence charge of the cation.
Covalent bonding delocalizes the $d$-electron charge and thus lowers the kinetic energy. The former mechanism favors the formation of local magnetic moments while the latter decreases the moment but increases the exchange coupling between the moments through virtual hopping processes.
In particular, the anion-mediated Kramers-Anderson ``superexchange'' between half-filled $3d$ orbitals often gives rise to strong antiferromagnetism.
Many $3d$ transition metal oxides can be classified as antiferromagnetic Mott insulators where the on-site Coulomb repulsion $U$ exceeds the electronic band width $W$.

$4d$ TMOs generally have significantly greater band widths and smaller $U$, due to the larger spatial extent of the $4d$ orbitals.
With $U$ and $W$ being more-or-less comparable in magnitude~\cite{hdkim,mravlje, wtian}, they are representative of the less-well understood intermediate coupling regime.
Without clear evidence of local moment formation and/or magnetic ordering, many of them are considered to be metallic Pauli paramagnets. The ruthenate family is a notable exception and features a variety of magnetically ordered phases. 
The best-known examples are the Ca-based Ca$_2$RuO$_4$ and Ca$_3$Ru$_2$O$_7$, and Sr-based Sr$_3$Ru$_2$O$_7$, Sr$_4$Ru$_3$O$_{10}$, and SrRuO$_3$ perovskites, featuring antiferromagnetic insulating and ferromagnetic metallic ground states, respectively~\cite{gcao}.
Their magnetic ordering temperatures are generally low, although recently SrRu$_2$O$_6$ has been reported to host high-temperature antiferromagnetism with a N\'eel temperature $T_{\rm N}=563$ K~\cite{hiley}.
Ruthenium dioxide (RuO$_2$), on the other hand, has been thought to fall in line with other binary $4d$ transition metal oxides ~\cite{mattheiss}; it is a good metal~\cite{schaefer} and believed to be Pauli paramagnetic~\cite{ryden}.
From the point of view of correlated electron physics and magnetism, RuO$_2$ seems to be one of the least interesting $4d$ TMOs. From a technology perspective, however, RuO$_2$ is by far one of the most important oxides.
It has numerous applications in electro- and heterogeneous catalysis, as electrode material in electrolytic cells, supercapacitors, batteries and fuel cells, and as diffusion barriers in microelectronic devices~\cite{over12}.
It owes its usefulness in part to its relatively high electrical conductivity combined with its excellent thermal and chemical stability~\cite{trasattibook}.
For the technological applications little attention has been paid to the potential role of magnetism (with the exception of Ref. \cite{torun}), presumably because magnetism is generally believed to be non-existent in bulk RuO$_2$.

In this letter we report on the finding that RuO$_2$ is distorted from the rutile symmetry ($P 4_2/m n m$) and exhibits antiferromagnetic order up to at least 300 K. Our DFT+$U$ calculations show that for a reasonable range of local interactions, the moments within the Ru$_2$O$_4$ rutile unit cell strongly prefer to align antiferromagnetically. The predicted magnetic order is confirmed with polarized neutron scattering experiments that show structurally forbidden peaks with a significantly decreased non-spin-flip/spin-flip intensity ratio. We conjecture that the relatively high N\'eel temperature can be attributed to the existence of half-filled $t_{2g}$ orbitals, in conjunction with a fairly large band width, similar to the recently reported case of SrTcO$_3$. Both materials can be described as strongly covalent intermediate coupling systems. An important distinction, however, is that RuO$_2$ is metallic and that its magnetism may be traced to a Fermi surface instability, whereas SrTcO$_3$ has been predicted to be a narrow gap insulator. These findings not only provide new insights into the origins of antiferromagnetism in the intermediate coupling regime, but may also have important ramifications for the understanding of the remarkable properties that make RuO$_2$ attractive for technological applications.

We begin our investigation of magnetism in RuO$_2$ with a DFT analysis (see ~\cite{sup} for technical details).
The majority of theoretical investigations considered bulk RuO$_2$ to be non-magnetic.
However, almost none of these studies considered the effects of on-site Coulomb interactions among the Ru 4$d$ orbitals.
Although these interactions are expected to be weaker than those in $3d$ TMOs, they should not be ignored.
Indeed, Ru $L_{23}$ X-ray-absorption spectra in combination with crystal-field-multiplet calculations indicated the importance of Coulomb interactions in RuO$_2$ ~\cite{zhu}.
Ru $3d$ core-level X-ray photoemission spectroscopy on RuO$_2$ was found to be most consistent with a dynamical mean field theory treatment of the single band Hubbard model, when $U$ is taken to be 1.8 eV compared to a bandwidth $W$ of 3.6 eV~\cite{hdkim}. Since interactions always play a critical role in magnetism it is imperative that we include their effects in our theoretical investigation.
To this end we employ the PBE+$U$ functional.

\begin{figure}[htp]
\includegraphics[width=1\columnwidth,clip=true]{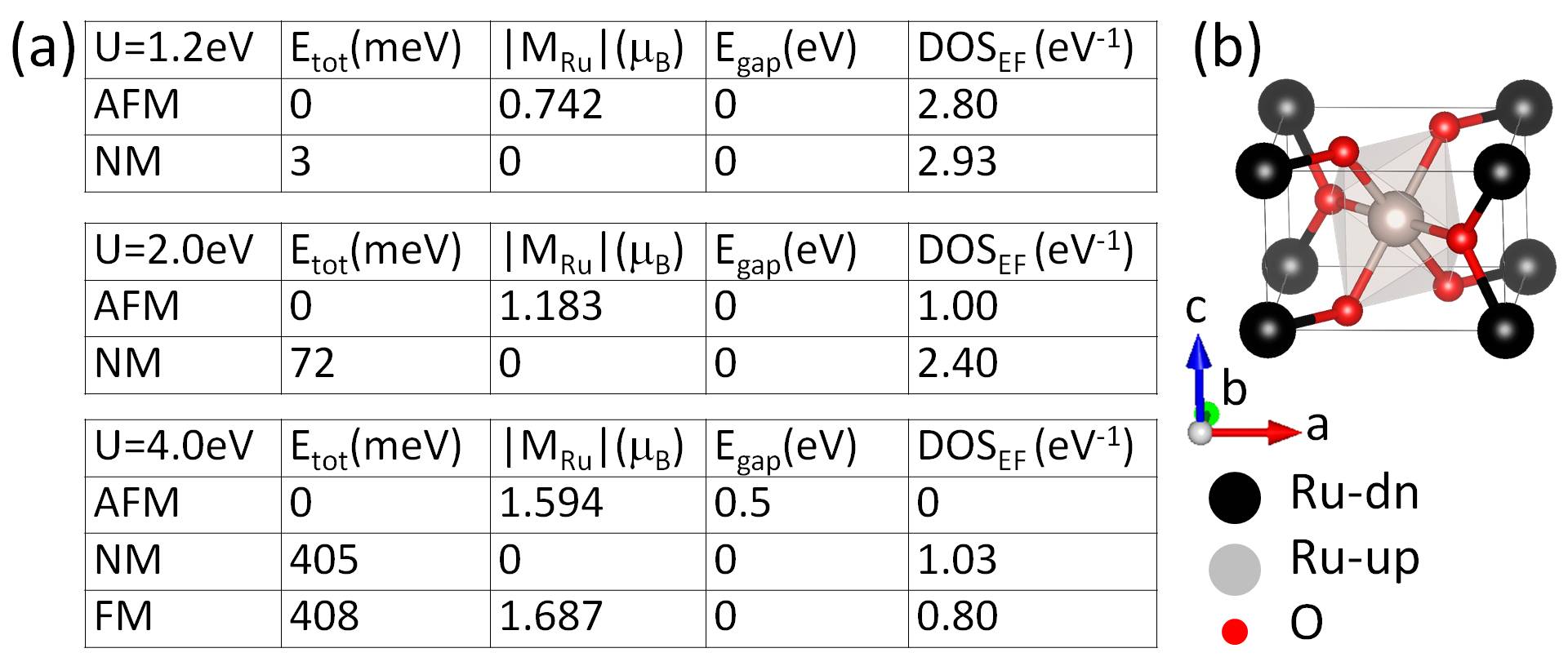}
\caption{\label{fig:fig1}
(color online)
(a) Total energy per Ru in meV, Ru magnetic moments within the rutile unit cell in $\mu_{\mathrm{B}}$, band gap in eV and density of states at the Fermi energy per eV per Ru, of the anti-ferromagnetic (AFM), the non-magnetic (NM) and the ferromagnetic (FM) configurations calculated with PBE+$U$
(b) Atomic and antiferromagnetic structure of bulk rutile RuO$_2$ as predicted by DFT and confirmed by neutron diffraction.
}\end{figure}

The DFT results are summarized in Fig. \ref{fig:fig1}(a).
First, we find that even for a weak $U$ of $1.2$ eV, the Ru moments within the rutile unit cell prefer to align antiferromagnetically (see Fig.\ref{fig:fig1}(b)).
With increasing $U$, the energy of the AFM configuration decreases relative to that of the non-magnetic structure. For $U=4$ eV, the system is no longer metallic and exhibits a band gap gap of about $0.5$ eV, contradicting the experimental fact that RuO$_2$ is a metal~\cite{schaefer}.
For $U=2$ eV we find that the AFM configuration is more stable than the non-magnetic one by a significant $72$ meV per Ru atom, while still retaining a sizable density of states at the Fermi level of about 1 eV$^{-1}$ per Ru atom.
The AFM configuration was reproduced using hybrid functionals.~\cite{yping, sup}
Apparently it requires an intermediate range of interaction strengths in RuO$_2$ to be simultaneously AFM and metallic.
We also considered the influence of the spin-orbit coupling, which we found to have a small effect on the electronic band structure and to only make small quantitative changes in the relative total energies of the magnetic configurations.~\cite{sup}
Ferromagnetic configurations turned out to be unstable or high in energy.

\begin{figure}[ht]
\includegraphics[width=1\columnwidth,clip=true]{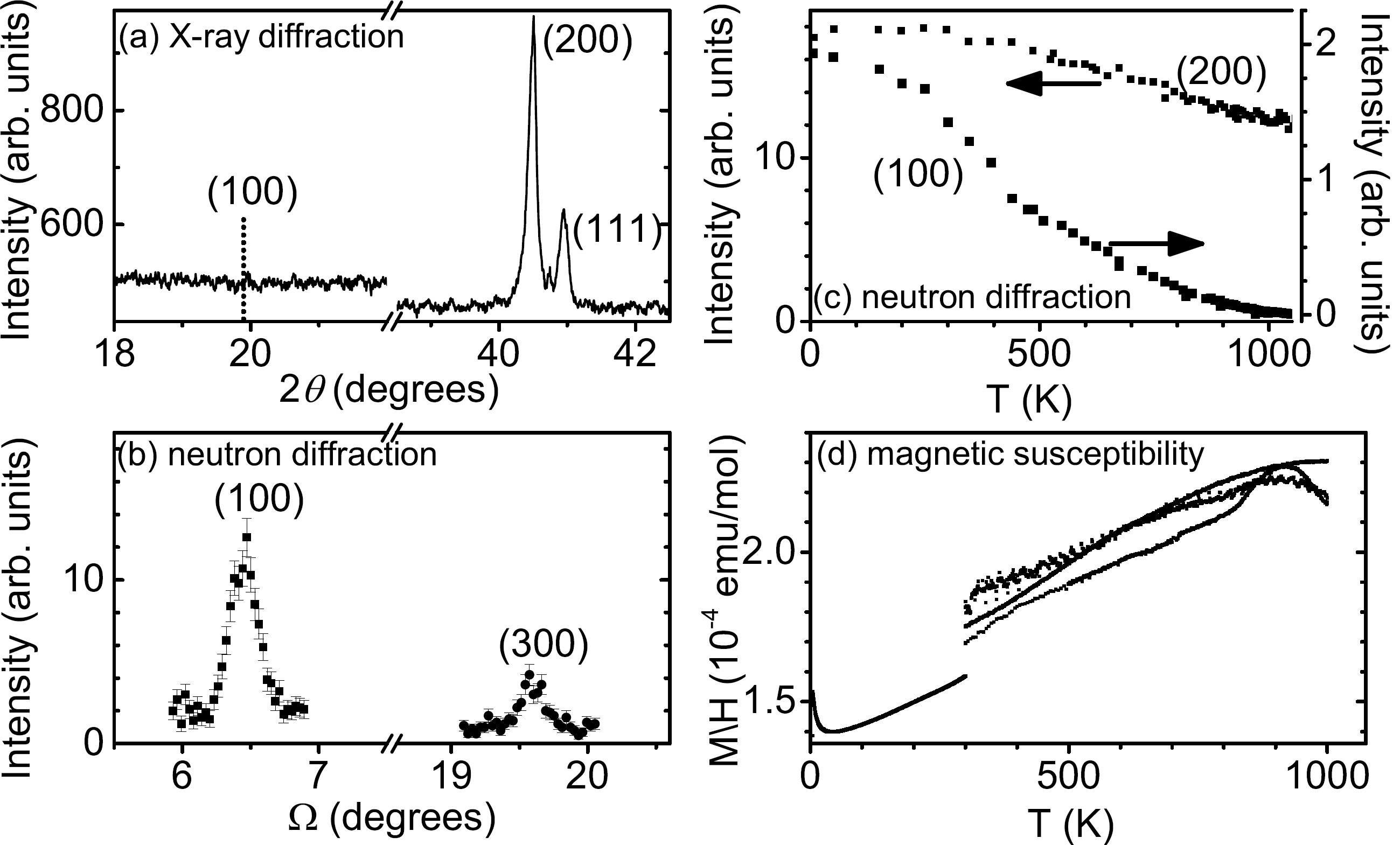}
\caption{\label{fig:fig2}
(a) X-ray and (b) unpolarized neutron diffraction data taken at 295 K at HB-3A.
(c) Temperature evolution of the integrated intensity of the nuclear (200) (left) and magnetic (100) (right) peak measured at HB-3A and HB-3.
(d) Magnetic susceptibility of different multigrain RuO$_2$ samples as a function of temperature.
}\end{figure}

To validate the DFT results, we synthesized RuO$_2$ single crystals via vapor transport in flowing oxygen, and subjected those crystals to extensive x-ray diffraction (XRD), neutron scattering, and magnetic susceptibility investigations \cite{sup}. For the perfect rutile structure, non-magnetic Ru contributions to the ($hkl$) Bragg reflection vanish when $h+k+l={\rm odd}$. The non-magnetic oxygen contributions vanish when $h+l={\rm odd}$ and $k=0$, or when $k+l={\rm odd}$ and $h=0$. Indeed, the XRD data in Fig. \ref{fig:fig2}(a), acquired at room temperature, show that the (100) Bragg peak is absent while the (200) and (111) peaks are clearly visible. On the other hand, room temperature unpolarized neutron diffraction data on a sample from the same crystal batch (Fig. \ref{fig:fig2}(b)) clearly reveal significant scattering intensity at reciprocal lattice points with odd indices such as (100) and (300), but not at the (001)and (003) locations. This would be consistent with the AFM configuration found from DFT, but it could also imply the existence of a lattice distortion that would be invisible when using a conventional x-ray source. In particular, the x-ray scattering cross section for light elements such as oxygen is very small. Interestingly, a polarized neutron scattering analysis of the (100) peak at 300 K \cite{sup}, indicates that RuO$_2$ is both distorted and antiferromagnetic at room temperature. While the majority contribution to the (100) peak intensity seems to be structural in origin, it does contain a magnetic scattering contribution: the non spin flip/ spin flip intensity ratio $R$ for the (200) peak is 12.8(2), whereas $R$ for the (100) peak is 8.0(2)~\cite{sup}. While this magnetic scattering contribution equates to only a small moment of about 0.05 $\mu_B$  at room temperature \cite{sup}, the presence of this magnetic moment is unambiguously demonstrated by the 60\% larger $R$ of the (200) peak as compared to the (100) peak. Given that at 300K the (100) peak intensity is close to being saturated (c.f. Fig. \ref{fig:fig2}(c)), a significant increase of the moment towards lower temperatures is unlikely.

The existence of room temperature antiferromagnetism is thus clearly established. However, the nature of the small lattice distortion is not understood. A symmetry analysis shows \cite{sup} that there are only two possible tetragonal subgroups of the rutile space group that could produce finite intensities for the forbidden reflections like (100) and (300).
Yet, a full refinement of the unpolarized neutron diffraction data involving over one hundred reflections, clearly converges to the rutile symmetry and, consequently, overestimates the magnetic moments.~\cite{sup}
Attempts within DFT+$U$ to find another total-energy minimum by breaking the rutile symmetry were unsuccessful~\cite{sup}.
At this point we are therefore not able to capture the nature of the distortion with a model that is consistent with the unpolarized neutron and X-ray scattering experiments, or the DFT+$U$ simulations, and leave this question for future investigations. The absence of the (001) reflection in neutron scattering implies the lack of a structural deformation along the c-direction and alignment of the magnetic moments along the rutile $c$-axis (the unpolarized neutron cross section vanishes when the scattering vector is parallel to the magnetic moment).
We note that the experimental magnetic moment of $\sim$0.05 $\mu_B$ from polarized neutron scattering is much smaller than the one predicted by DFT. Such discrepancies between DFT and experiment are quite common in metallic antiferromagnets, such as for example the Fe based superconductors ~\cite{nmannella,yttam}, and probably reflect the inability of the static mean field DFT to capture charge and spin fluctuations in time and space.

Fig. \ref{fig:fig2}(c) shows the full temperature dependence of the (100) and (200) diffraction intensities. The (100) peak vanishes near 1000 K while the (200) peak persists to higher temperature and diminishes in intensity according to the Debye-Waller factor. This rules out multiple scattering as the origin of the (100) reflection, because the temperature dependences of the (100) and (200) peaks are clearly different. The concave temperature dependence of the (100) peak intensity furthermore suggests that the magnetic and/or structural ordering is fairly short-range. This is consistent with the Lorentzian lineshape of the (100) peak, as opposed to the Gaussian lineshape of the (200) reflection \cite{sup}. 

The presence of room temperature antiferromagnetism goes against the current lore that RuO$_2$ is a Pauli paramagnet. This general belief probably stems from the early work by Ryden {\it et al} ~\cite{ryden} that concluded Pauli paramagnetism from the quadratic temperature dependence of the magnetic susceptibility within 4-300 K. However, older measurements of the magnetic susceptibility ~\cite{guthrie, fletcher} were performed for much larger temperature ranges up to 1000 K and demonstrated instead a linear increase as a function of temperature. We repeated those measurements while ramping the temperature continuously from 4 K to 300 K and from 300 K to 1000 K. The results are presented in Fig. \ref{fig:fig2}(d). The value of $1.7\times 10^{-4}$ emu/mol ($300$ K) is in good agreement with previous reports ~\cite{guthrie, fletcher, ryden}. The 30\% rise of the magnetic susceptibility with increased temperature from 300 K to 1000 K is also in excellent agreement with Fletcher et al.~\cite{fletcher}, the only study that measured up to 1000 K. Our measurements, however, either produce a clear, broad maximum in the susceptibility or a significant leveling at the highest temperature, consistent with the presence of short-range ordering. Due to the extreme difficulty in measuring small magnetic signals at such high temperature, which is near the limit of our instrument capability, as well as the possible loss of oxygen, different crystals produce slightly different behavior above 850 K. It is possible that this changing magnetic behavior above 900 K is related to the vanishing of the (100) peak and its underlying magnetic and/or structural order.

\begin{figure}
\includegraphics[width=1\columnwidth,clip=true]{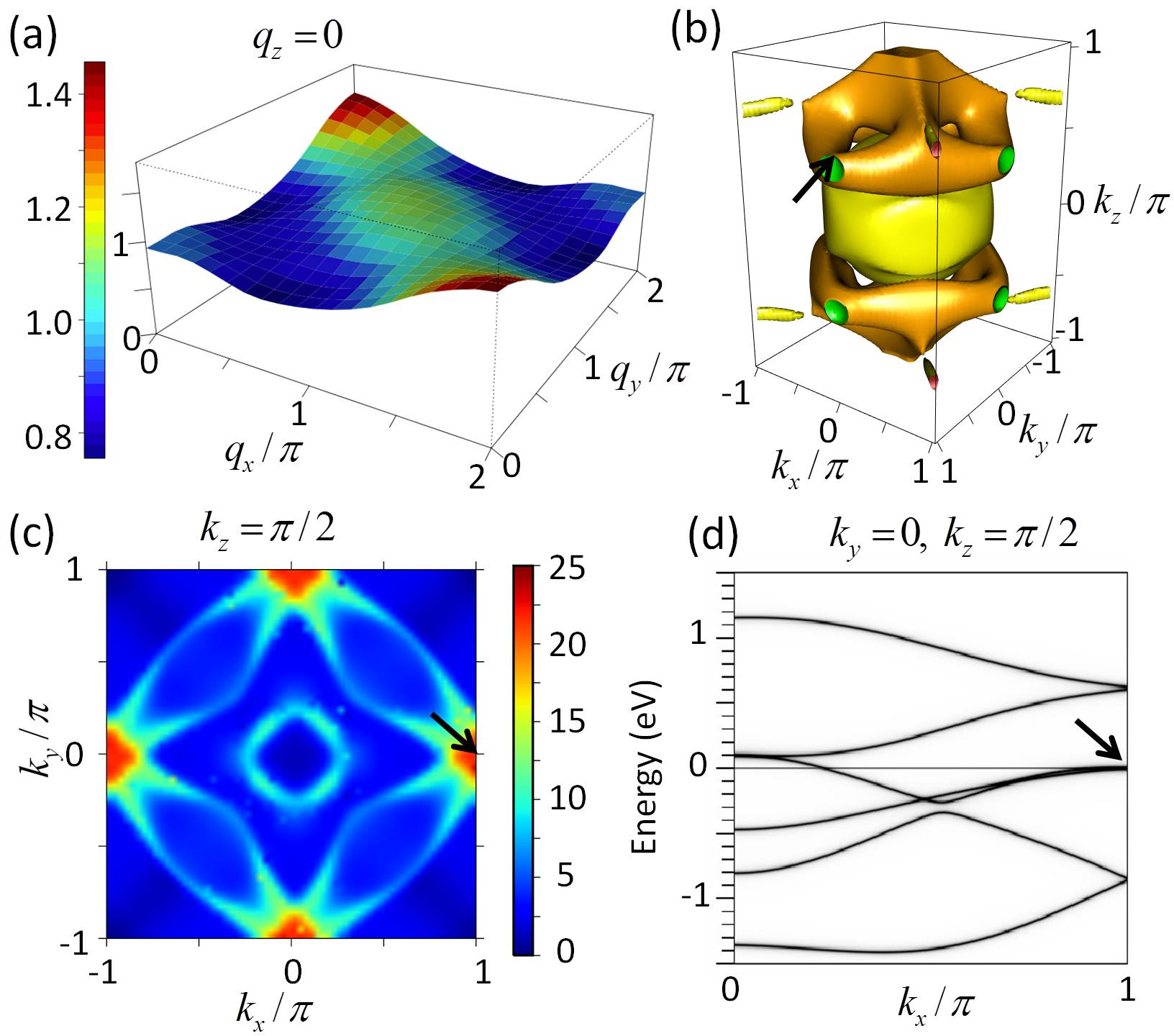}
\caption{\label{fig:fig3}
(color online)
(a) Lindhard response function $\chi_0(q)$ in eV$^{-1}$.
(b) Fermi surface.
(c) Momentum resolved contribution to $\chi_0(q=(2\pi,0,0))$.
(d) Band structure at $k_y=0$ and $k_z=0.5\pi$ along the $k_x$ direction.
Black arrow in various panels indicates one of eight nested ``hot spots''.
}\end{figure}

Given the itinerant nature of the conduction electrons in RuO$_2$, antiferromagnetism possibly originates from a spin density wave instability of the Fermi surface.  To explore this possibility, we calculate the Lindhard response function. To this end, we first map the first-principles electronic structure from a non-magnetic DFT calculation (with $U=0$) onto a low energy effective model that only contains the Ru $4d$ orbital degrees of freedom. Specifically we perform a Wannier transformation of the $2\times 5$ Ru $4d$ bands within the $[-2,6.5]$ eV energy window.
The resulting tight binding Hamiltonian then allows us to efficiently compute the Lindhard response $\chi_0(q)$ as a function of the momentum $q$ according to:
\begin{eqnarray}\label{chi}
\chi_0(q)&=&\sum_k \Big\{ \sum_{st\mu\nu}\frac{\langle s|\mu k\rangle \langle \mu k|t\rangle
\langle t|\nu k+q\rangle \langle \nu k+q|s\rangle}{E_{\nu}(k+q)-E_{\mu}(k)}
\nonumber \\ && \big( f(E_{\nu}(k+q))-f(E_{\mu}(k))\big) \Big\}
\end{eqnarray}
where $s,t/\mu,\nu$ denote the orbital/band indices, respectively, $k,k+q$ the momenta, and $f(E)$ the Fermi distribution function at energy $E$.
As shown in Fig. \ref{fig:fig3}(a), the response function $\chi_0$ is peaked at $q=(2\pi,0,0)$ and $q=(0,2\pi,0)$ with a value of roughly 1.4 eV$^{-1}$.Therefore within the random phase approximation~\cite{graser} the interacting response function, given by $\chi = \chi_0/(1-U \chi0)$, will diverge for interactions larger than U $\approx$(1/1.4)eV) driving the system towards a spin density wave instability.

Such a spin density wave (or AFM) modulation would produce magnetic reflections at the (100) and (010) locations in reciprocal space, exactly as predicted by DFT and observed experimentally.
Fig. \ref{fig:fig3}(c) shows the crystal momentum resolved contribution to the magnetic susceptibility, obtained from evaluating the term in the curly bracket in equation (\ref{chi}) as a function of crystal momentum $k$ for the fixed momentum $q=(2\pi,0,0)$.
From the heat map in Fig. \ref{fig:fig3}(c) we see that the dominating contributions originate from states near $(\pi,0,\pi/2)$ and symmetry related points.
These ``hot spots'' are located at the neck-shaped regions of the Fermi surface as indicated by the black arrows in Fig. \ref{fig:fig3}(b).
The energy bands near these $k$-points have very low Fermi velocities (Fig. \ref{fig:fig3}(d)) and thus contribute a large density of states to the magnetic instability.
The hot spots are folded on top of one another via translation by a reciprocal lattice vector. This can be seen in Fig. \ref{fig:fig3}(d). Note that the bands at the hot spot location are doubly degenerate, which is a consequence of the rutile symmetry.
This degeneracy is lifted by antiferromagnetism and as a consequence of the hot spots being folded on top of each other, the magnetic unit cell equals the structural unit cell.
This hot spot mechanism differs from the classical example of chromium~\cite{efawcett} in which nesting takes place between large parallel sheets of Fermi surface, but is analogous to that being proposed for charge density waves in $2H$-NbSe$_2$ ~\cite{borisenko}.
Whether the Fermi surface hot spots in RuO$_2$ are capable of driving the antiferromagnetism or rather play an assisting role, remains an open question - one that is an integral part of the longstanding debate on itinerant versus localized magnetism in metallic systems~\cite{wohlfarth,moriya}.

The discovery of AFM in RuO$_2$ and particularly its relatively high N\'eel temperature ($\ge$300K) is significant because metallic AFM oxides are rare ~\cite{phobe,gzhang}, and their ordering temperatures are generally low. For example, within the $3d$ series, CaCrO$_3$ and Sr$_{0.99}$Ce$_{0.01}$MnO$_3$ have a $T_{\rm N}$ of 90 K~\cite{komarek} and 220 K~\cite{sakai}, respectively. In the $4d$ series, the ruthenates Ca$_3$Ru$_2$O$_7$  and Na-doped CaRuO$_3$ ($T_{\rm N}=70$ K) display antiferromagnetism at $T_{\rm N}=56$ and 70 K, respectively, significantly lower than that of RuO$_2$~\cite{gcaoca3ru2o7,mshepard}, and they are borderline insulating.
Indeed, the recent discoveries of AFM with high $T_{\rm N}$ in $4d$ transition metal oxides were made in semiconducting SrRu$_2$O$_6$ ($T_{\rm N}=563$ K)~\cite{hiley} and SrTcO$_3$ ($T_{\rm N}=1023$ K)~\cite{rodriguez}; the latter is theoretically predicted to be insulating. ~\cite{rodriguez,mravlje}
While the debate on itinerant versus localized magnetism in metallic systems~\cite{wohlfarth,moriya} shows that it is difficult to determine the role of itineracy in AFM order, the robust metallicity, small moment, and high magnetic ordering temperature of RuO$_2$ places it in a regime that was hitherto not accessible in transition metal oxides.

The relatively high $T_{\rm N}$ in RuO$_2$ appears to be consistent with recent explanations of high temperature AFM in SrTcO$_3$ ~\cite{mravlje} and SrRu$_2$O$_6$~\cite{wtian}.
Here it was argued that $T_{\rm N}$ maximizes in a regime in which the ratio of the interaction $U$ and the band width $W$ is large enough to form robust magnetic moments, but small enough to allow for significant exchange interactions between those moments.
Both high-$T_{\rm N}$ compounds share another important feature, namely the existence of a $4d^3$ electron configuration. Since at $T_{\rm N}$ SrTcO$_3$ has the ideal perovskite symmetry (space group \textit{Pm$\bar{3}$m})~\cite{rodriguez}, the three $t_{2\rm g}$ orbitals are degenerate and thus half-filled. In SrRu$_2$O$_6$ the RuO$_6$ octahedra are stretched along the {\textit c}-axis, but the $C_{3v}$ symmetry still protects the $t_{2g}$ orbital degeneracy~\cite{dwang}. Hence, the 3 $t_{2\rm g}$ orbitals are also half-filled.
A simple chemical bonding picture by Moriya ~\cite{moriya2} explains why antiferromagnetism (localized or itinerant) is particularly stable at half filling: the majority spin states on one magnetic sublattice hybridize with the minority spin states on the other sublattice, and the resulting ``chemical bond'' is most stable at half filling while the stabilization energy is determined by the band width.

At first sight, RuO$_2$ does not seem to match this picture as it has a $4d^4$ electron count.
However, our orbital resolved density of states shows \cite{sup}  that the $4d_{x^2-y^2}$ orbital is filled with two electrons and resides below $E_{\rm F}$, leaving the remaining two $t_{2g}$ orbitals half filled. 
Hence the specific crystal field splitting of the edge-shared octahedra in the rutile structure ensures that the $4d_{xz}$ and $4d_{yz}$ $t_{2g}$ orbitals that are relevant for the AFM order are formally half filled, similar to SrTcO$_3$ and SrRu$_2$O$_6$. An important distinction, however, is that RuO$_2$ is a good metal whereas SrTcO$_3$ is theoretically predicted to be insulating~\cite{rodriguez,mravlje} and SrRu$_2$O$_6$ has been determined to be semiconducting from resistivity measurements. ~\cite{hiley}
The unique combination of good metallicity and high temperature AFM in RuO$_2$ will allow for a more complete benchmarking of theoretical models describing the interplay between magnetism and metallicity in oxide materials.

Our discovery of antiferromagnetism in a strongly metallic binary oxide material also calls for the reevaluation of the magnetic properties of other 4d and 5d metallic oxide systems. Many of these materials already are of technological importance, often as catalyst or other chemical applications, but the existence of itinerant antiferromagnetism in this class of materials would open up a new realm of possibilities, specifically in light of recent developments in antiferromagnetic-metal spintronics.~\cite{bgpark} Here it may be needed to enhance the magnetic properties, such as the moment, via e.g. alloy substitution or dimensional confinement.

TB and PCS contributed equally to this work.
We thank Veerle Keppens for the use of her laboratory equipment.
The research was supported by the U.S. Department of Energy, Office of Science, Basic Energy Sciences, Materials Sciences and Engineering Division (TB, PCS, OD, YW, PRCK, HHW). Work by TAM (response function calculation) was performed at the Center for Nanophase Materials Sciences, a DOE Office of Science user facility.
HDZ (crystal growth, XRD and low temperature susceptibility measurements) acknowledges support from NSF-DMR-1350002.
MRK (high temperature susceptibility measurements) acknowledges support from the Gordon and Betty Moore Foundations EPiQS Initiative through Grant GBMF4416. Research at ORNL's High Flux Isotope Reactor (HBC, MM, SXC) was sponsored by the Scientific User Facilities Division, Office of Basic Energy Sciences, US Department of Energy. This research used resources of the National Energy Research Scientific Computing Center, a DOE Office of Science User Facility supported by the Office of Science of the U.S. DOE under Contract No. DE-AC02-05CH11231.

\begin{widetext}
{\Large\bf Supplemental Material}

\section{Crystal growth}

Crystals of RuO$_2$ with the rutile structure were obtained using vapor transport growth in a quartz tube.~\cite{yshuang}
The starting material ($>99.95\%$ anhydrous RuO$_2$ powder) was heated to 1350 $^{\circ}$C in flowing oxygen,
producing RuO$_3$ gas that was allowed to condense into RuO$_2$ at the cold end of the tube at 1150 $^{\circ}$C.
Crystals with sizes ranging from 0.5 to 1.5 mm were obtained.

\section{X-Ray Diffraction}

\begin{figure}[htp]
\includegraphics[width=0.6\columnwidth]{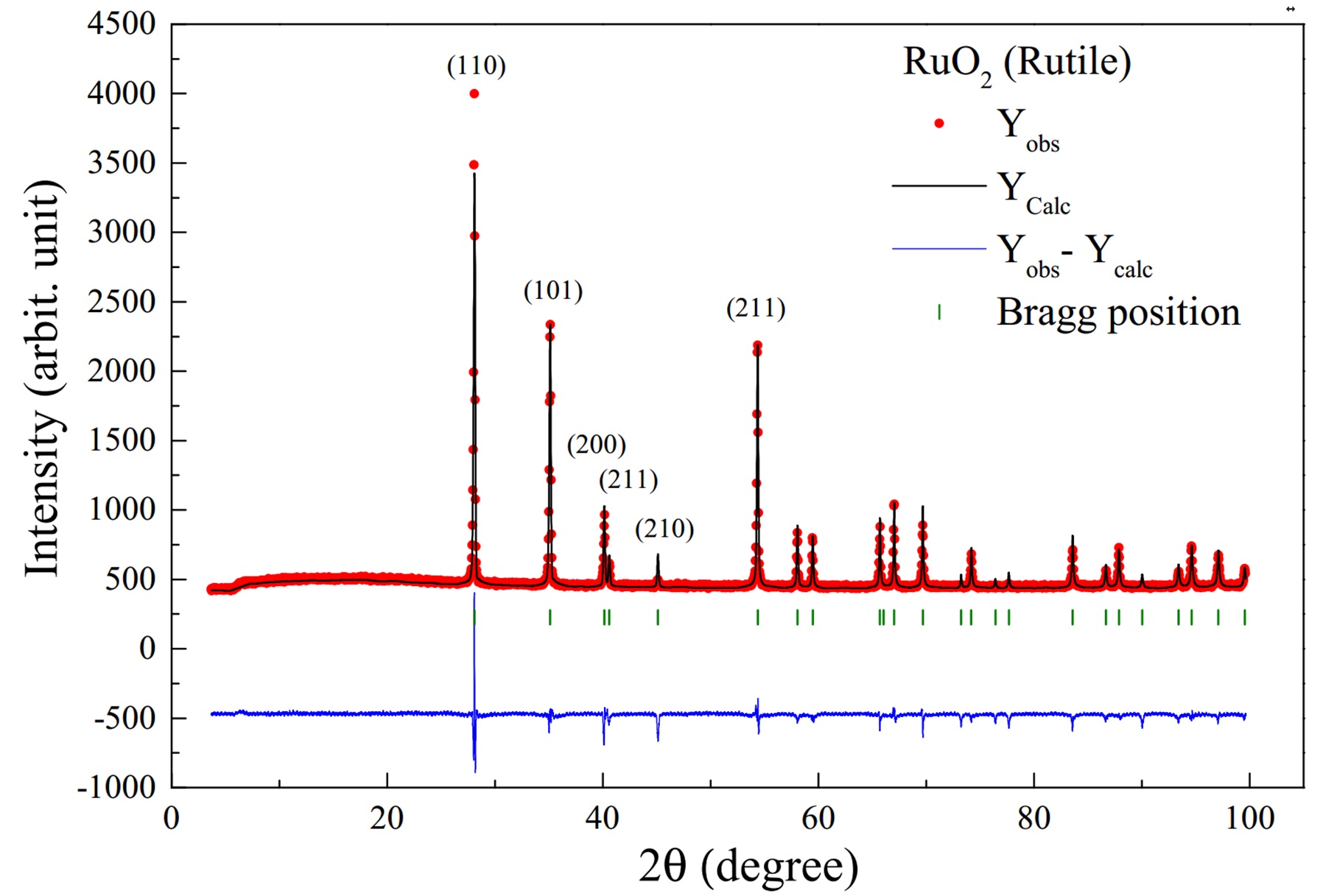}
\caption{\label{fig:figsup1}
Measured (Y\_{obs}) and calculated (Y\_{calc}) X-Ray diffraction data, their difference, as well as the expected Bragg reflection positions for a perfect rutile lattice.
}
\end{figure}

Room temperature X-ray diffraction (XRD) patterns on crushed single crystals were measured with a HUBER X-ray powder diffractometer. Structural refinements were performed using the software package FullProf Suite \cite{fullprof}. All peaks can be indexed with a rutile structure with space group $P 4_2/m n m$. The mismatch in the Bragg peak intensities is due to the preferred orientation of grains, which is common for crushed crystal samples measured in the flat-plate geometry.

\section{Neutron diffraction}

Single crystal unpolarized neutron diffraction was performed at the HB-3A Four-Circle Diffractometer (FCD) equipped with a 2D detector at the High Flux Isotope Reactor (HFIR) at ORNL. A neutron wavelength of 1.005 \AA~from a bent perfect Si-331 monochromator \cite{hb3a} was used for data collection. The refinement was performed by the FullProf Suite \cite{fullprof}. A total of 118 reflections was measured by rocking curve scans at 295 K; 89 symmetry allowed reflections were used for structure refinement within the space group $P 4_2/m n m$. The structural parameters of  RuO$_2$, measured at 295~K, are listed in Table~\ref{strs}. Given that the unit cell is tetragonal, a symmetry analysis shows that there are only two possible subgroups (space group 77 and 81) that could show forbidden peaks at the (100) and (300) locations in reciprocal space~\cite{inttabcrys}.
Yet, when using these lower subgroup symmetries, the forbidden reflections still could not be fitted to the experimental data.
Neutrons diffraction intensities were checked at $(\frac{1}{2}\frac{1}{2}\frac{1}{2})$ and $(\frac{1}{2}10)$, but no peak signals were observed here. 
A G-type antiferromagnetic structure on the Ru sublattice with the magnetic moments pointing along the $c$-axis can fit the observed (100) and (300) reflections, and is consistent with the (001) reflection being absent. The overestimated magnetic moment extracted from this analysis is 0.23 $\mu_B$ per Ru atom.
\\
\\
\\
\\
\begin{table}[htp]
\caption{The structure parameters of RuO$_2$ measured at 295~K by single crystal neutron diffraction at HB-3A. The space group is $P 4_2/m n m$, $a$=$b$= 4.492(2) \AA, $c$=3.1061(15) \AA, $\alpha$=$\beta$=$\gamma$=90$^o$. $R_f$=0.0231. $\chi^2$=4.31.  The atomic displacment parameters $U$ have units of \AA$^2$.}
\begin{tabular}{c|c|c|c|c|c|c}
\hline
atom & $type$ &$site$& $x$& $y$ &  $z$  & $U$\\
\hline
Ru1 & Ru & $2a$&0   &0  &0    & 0.0021(4) \\
O1 & O & $4f$ & 0.30558(17)  & 0.30558(17)  &0 &  0.0040(4) \\
\hline
\end{tabular}
\label{strs}
\end{table}
\\
\\
\begin{figure}[htp]
\includegraphics[width=1\columnwidth]{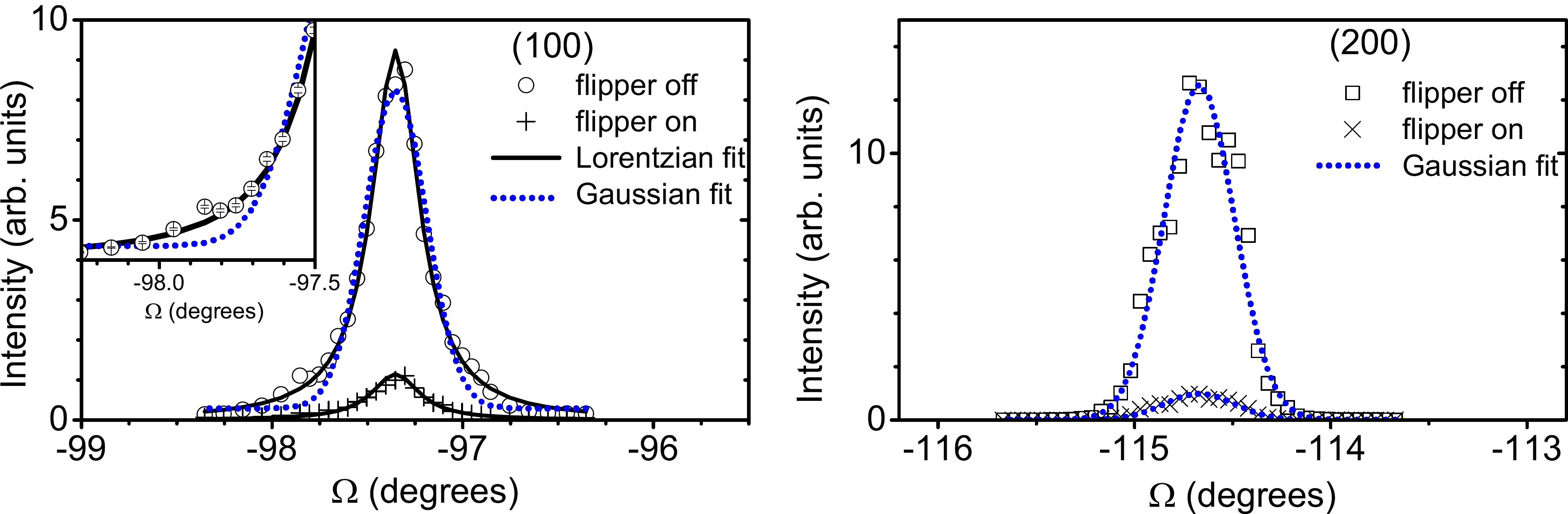}
\caption{\label{fig:figsup1b}
The (100) (left) and (200) (right) Bragg reflections measured using polarized neutron diffraction with the flipper on and off measured at HB-1. The peaks  measured with the flipper off are fit by both a Gaussian (solid line) and a Lorentzian (dotted line for the (100) peak, and by a Gaussian for the (200) peak. The inset in the left panel shows a magnification of the tail of the (100) peak measured with the flipper off, revealing that a Lorentzian function fits the (100) lineshape better than a Gaussian function.
}
\end{figure}
\\
\\
Polarized neutron diffraction experiments were carried out on a thermal neutron triple-axis spectrometer HB-1 installed at the High Flux Isotope Reactor (HFIR) at ORNL. Heusler alloy (111) crystals were used as monochromator and analyzer. Neutrons with an incident energy of 13.5 meV were used, together with a horizontal collimator sequence of $48'$--$80'$--S--$80'$--$240'$. The energy resolution is $\sim$1.5 meV. Contamination from higher-order beams was effectively eliminated using PG filters. While the polarized and unpolarized neutron data were recorded from different samples on different beam lines, the ratio of the (100) and (200) peak intensities were consistent.

To extract the integrated intensities of the measured (200) and (100) Bragg reflections, we attempted to fit the peaks with a Gaussian and a Lorentzian lineshape. In the left panel of Fig. \ref{fig:figsup1b}, the results are shown for the (100) peak. For the flipper off peak (open symbols), the dotted line is a Gaussian fit and the solid line is a Lorentzian fit. It is clear that the measured peak shape contains tails that are not captured by the Gaussian fit. This is also highlighted in the inset of the left panel, where the error bars of the measured intensity are also included; they are smaller than the size of the symbols used for the datapoints. Based on these findings we used a Lorentzian fit to extract the integrated intensity of the measured (100) peak. The (200) peak instead is fit well with a Gaussian curve, as shown in the right panel of Fig. \ref{fig:figsup1b}. Note that the Lorentzian shape of the (100) peak suggests a glassy or short-range ordered state.

Using these fits, a reference flipping ratio of 12.8(2) was measured at the nuclear (200) reflection, corresponding to an 86\%\ polarization of the beam.
In order to determine the magnetic contribution to the (100) reflection, we determined the intensities in the flipper on and off channels with the neutron spin parallel to the $Q$ direction.
The ratio of the flipper on and off intensities ($I\rm_{off}$/$I\rm_{on}$) is 8.0(2). Considering the instrumental flipping ratio of 12.8(2), the magnetic fraction at the (100) reflection is estimated to be about 5\%.
If instead the poorly fitting Gaussian function is used to fit the (100) peak, the ratio $I\rm_{off}$/$I\rm_{on}$ becomes 9.3(2).
\\
\\
If we apply the polarized neutron result, that estimates a 5\% magnetic scattering fraction at the (100) reflection, to the unpolarized neutron diffraction data refinement that resulted in a perfect rutile structure, then the ordered magnetic moment will be reduced to from 0.23 $\mu_B$ to 0.05 $\mu_B$ (0.23x$\sqrt{0.05}$$\mu_B$). The remaining 95\% intensity of the (100) reflection is likely caused by additional structural distortions and a larger unit cell is likely required to fit those reflections.  The small number of forbidden reflections did not allow us to solve the distorted structure using a larger unit cell.

\section{Magnetic susceptibility}

The dc magnetic susceptibility measurements were performed on as-grown samples using a magnetic field of 2 T while heating at a rate of 2 K/minute. The low temperature data were measured using a Quantum Design superconducting interference device (SQUID) magnetometer on multiple grains that were randomly oriented, having a total mass of 89.8 mg.

The high temperature data were measured using a Quantum Design VersaLab with the high temperature VSM option on multiple crystals that were randomly oriented with a total mass varying from 11 mg to 42 mg, and on a collection of rod-like RuO$_2$ crystals that were aligned with the rutile c-axis parallel to the magnetic field.
 
\section{First principles calculations}\label{sec:secdft}

\begin{figure}[htp]
\includegraphics[width=0.6\columnwidth]{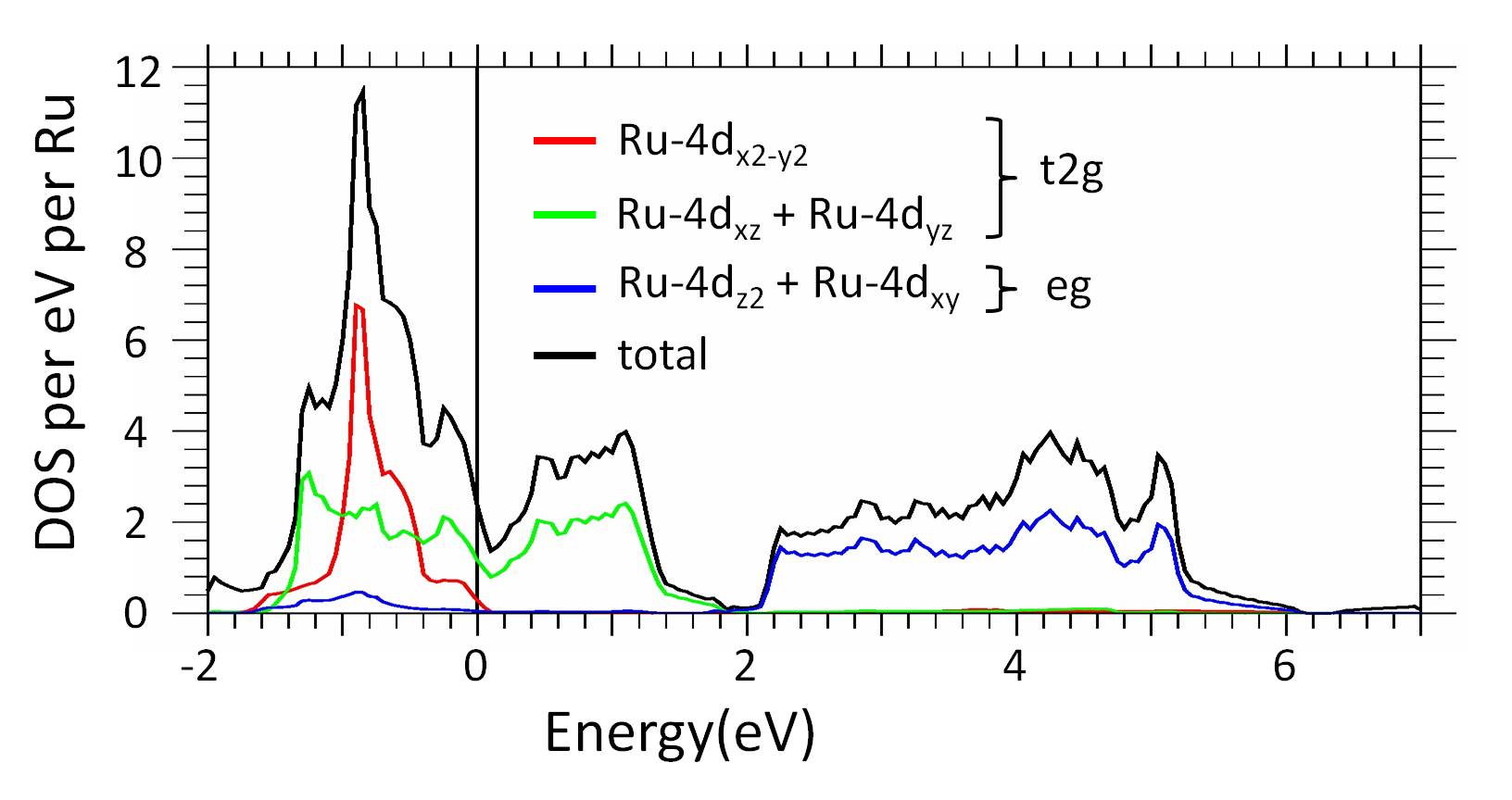}
\caption{\label{fig:figsup2}
Orbital resolved density of states from the non-magnetic PBE+$U$ simulation ($U$=2eV).
}
\end{figure}

The Density Functional Theory (DFT) calculations are performed using the plane wave projector augmented wave (PAW) method~\cite{paw} as implemented in the VASP code \cite{vasp1,vasp2,vasp3}.
For the PBE+$U$ calculations the double counting correction scheme by Dudarev \textit{et al.}\cite{dudarev} and a $8\times8\times12$ k-mesh is employed. Atomic positions and lattice constants are converged down to 1 meV/\AA. The density of states calculations of are performed on a $16\times16\times24$ k-mesh. 
The VASP calculations have been performed without the use of symmetry by using the flag ISYM=0 in the input file.
This is done because there appears to be a bug in the symmetry routine of VASP.
Only when the symmetry is switched off do we obtain equal moments on the symmetry equivalent Ru atoms.
The number of plane waves is determined by an energy cut-off of 500 eV.  The Wannier calculations where performed with Wannier90~\cite{wannier90}. The number of maximal localization steps was set to zero in order to obtain symmetry respecting projected Wannier functions~\cite{ku}.
\\
\\
The orbital resolved density of states is shown in Fig. \ref{fig:figsup2}.
It shows that the $4d_{x^2-y^2}$ orbital is fully filled.
This result is quite insensitive to the value of U being used. The filling of the $4d_{x^2-y^2}$ band varies from 97\% to 98\% when using U=0 and U=2eV respectively. The $4d_{x^2-y^2}$ being filled leaves the remaining two $t_{2g}$ orbitals, $4d_{xz}$ and $4d_{yz}$, half filled. 
Here we follow the convention of Ref. ~\cite{veyert}: the $4d_{x^2-y^2}$ orbital has its lobes pointing in the $(001)$, $(1\overline{1}0)$ and $(\overline{1}10)$ directions.

\begin{table}[htp]
\includegraphics[width=0.8\columnwidth]{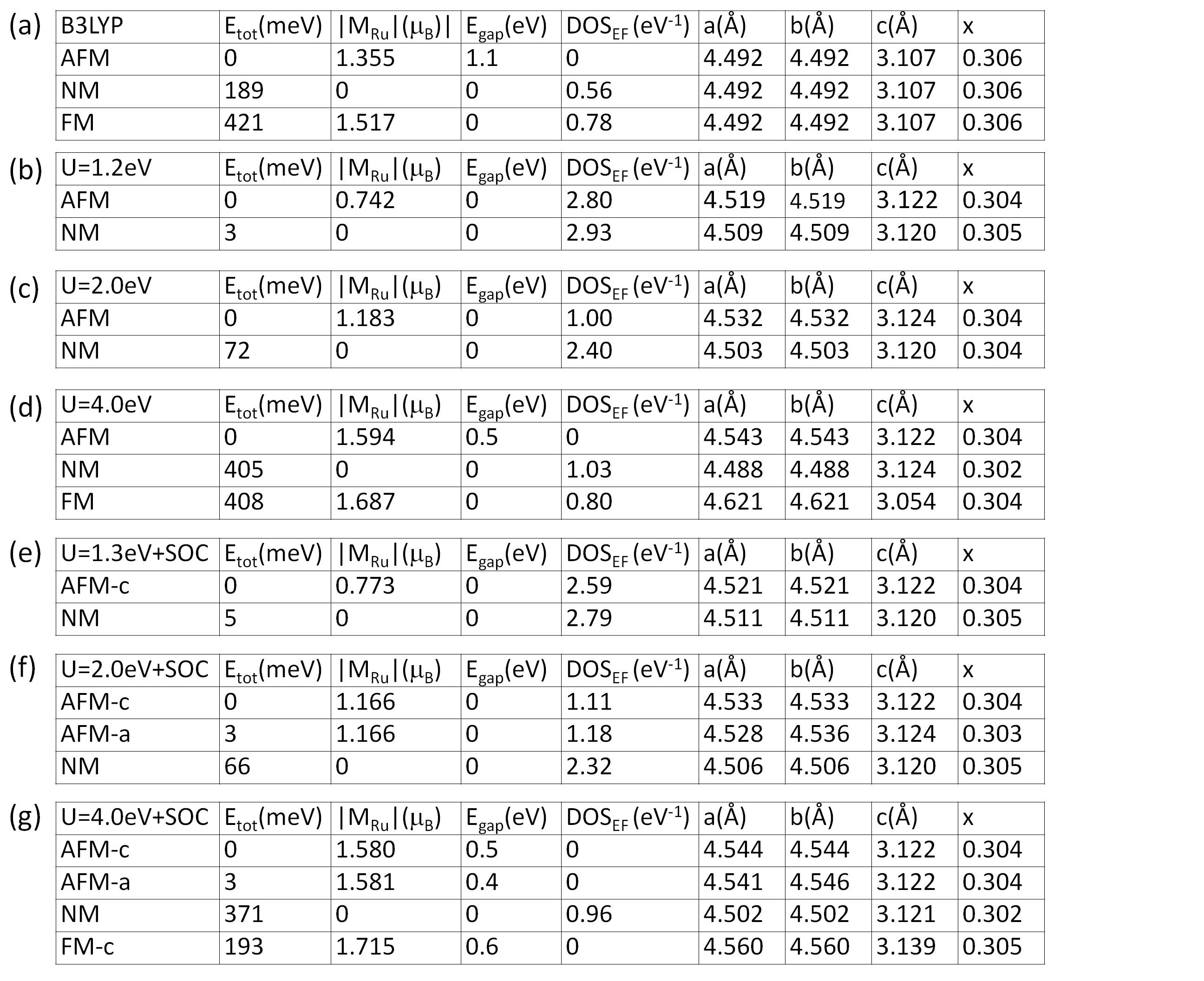}
\caption{\label{fig:figsup3}
Total energy per Ru in meV, value of the Ru magnetic moments
within the rutile unit cell in $\mu_{\mathrm{B}}$, band gap in eV, density of states at the Fermi energy per eV per Ru, lattice constants in \AA ~and internal lattice parameter $x$ that fixes the oxygen positions for the non-magnetic (NM), the anti-ferromagnetic (AFM) and ferromagnetic (FM) configurations with the moments along the $a$ and $c$ crystallographic axes calculated with B3LYP and PBE+$U$ without and with spin orbit coupling (SOC).
}
\end{table}

 Hybrid functional calculations have been performed using B3LYP with an energy cut-off of 400 eV and a $4\times4\times6$ k-mesh
due to the computationally demanding evaluation of the Fock matrix (especially without the use of symmetry as mentioned above). For the same reason the lattice structure in the B3LYP calculations was taken from x-ray diffraction reported in Ref. ~\cite{boman} without relaxing the atomic positions or the lattice constants. 
Table \ref{fig:figsup3}(a) shows that the AFM configuration is reproduced using the B3LYP hybrid functional.
However, this functional produces a gap of 1.1 eV, even larger than that of the PBE+$U$ calculation with $U$=4 eV.
The same AFM configuration has also been reported in Ref.~\cite{yping} from a calculation with the B3PW91 hybrid functional but no mention was made whether this configuration was still properly metallic.
The B3LYP hybrid functional used here has the same parametrization as the B3PW91 functional used in Ref.~\cite{yping}.
In particular both functionals~\cite{pjstephens,abecke} contain 20\% of Fock exchange.
\\
\\
To investigative the possibility of a structural distortion we removed the symmetry by moving the atoms 0.4 \AA ~in random directions both in the Ru$_2$O$_4$ unit cell and in a $2\times2\times2$ supercell. Upon minimizing the forces we found the atoms to relax back to the positions dictated by rutile symmetry, both in the non-magnetic $U$=0 and the anti-ferromagnetic $U$=2 eV simulation. We also note that a phonon calculation reported in Ref. ~\cite{bohnen} did not show phonons with imaginary frequencies.
\\
\\
From comparing Table \ref{fig:figsup3}(c) and Table \ref{fig:figsup3}(f) we see that for the $U$=2 eV simulations the spin-orbit coupling induces no qualitative changes. The same is the case for the AFM and NM $U$=4 eV simulations, but for the FM $U$=4 eV simulations the SOC does seem to have a strong effect (c.f. Table \ref{fig:figsup3}(d)(g)).
The minimum $U$ for which the AFM state can be stabilized slightly increases from $U$=1.2eV to $U$=1.3eV upon including SOC in the simulations (c.f. Table \ref{fig:figsup3}(b)(e)).
The effects of the spin-orbit coupling on the non-magnetic ($U$=0) band structure, used as input for the Lindhard response calculations in the manuscript, is shown in the band structure close to the Fermi-level presented in Fig. \ref{fig:figsup4}.
We note that upon including SOC in our DFT calculation, the band degeneracy at the ``hotspot'' location is lifted. This could explain why in the DFT simulations the AFM ground state energy (relative to the NM ground state energy) increases upon including SOC as the SOC term will partly lift the degeneracy of the ``hotspots'' that contribute to the AFM instability.
It will be interesting in future investigations to quantify the effects of the SOC on the Lindhard response function.

\begin{figure}[htp]
\includegraphics[width=0.6\columnwidth]{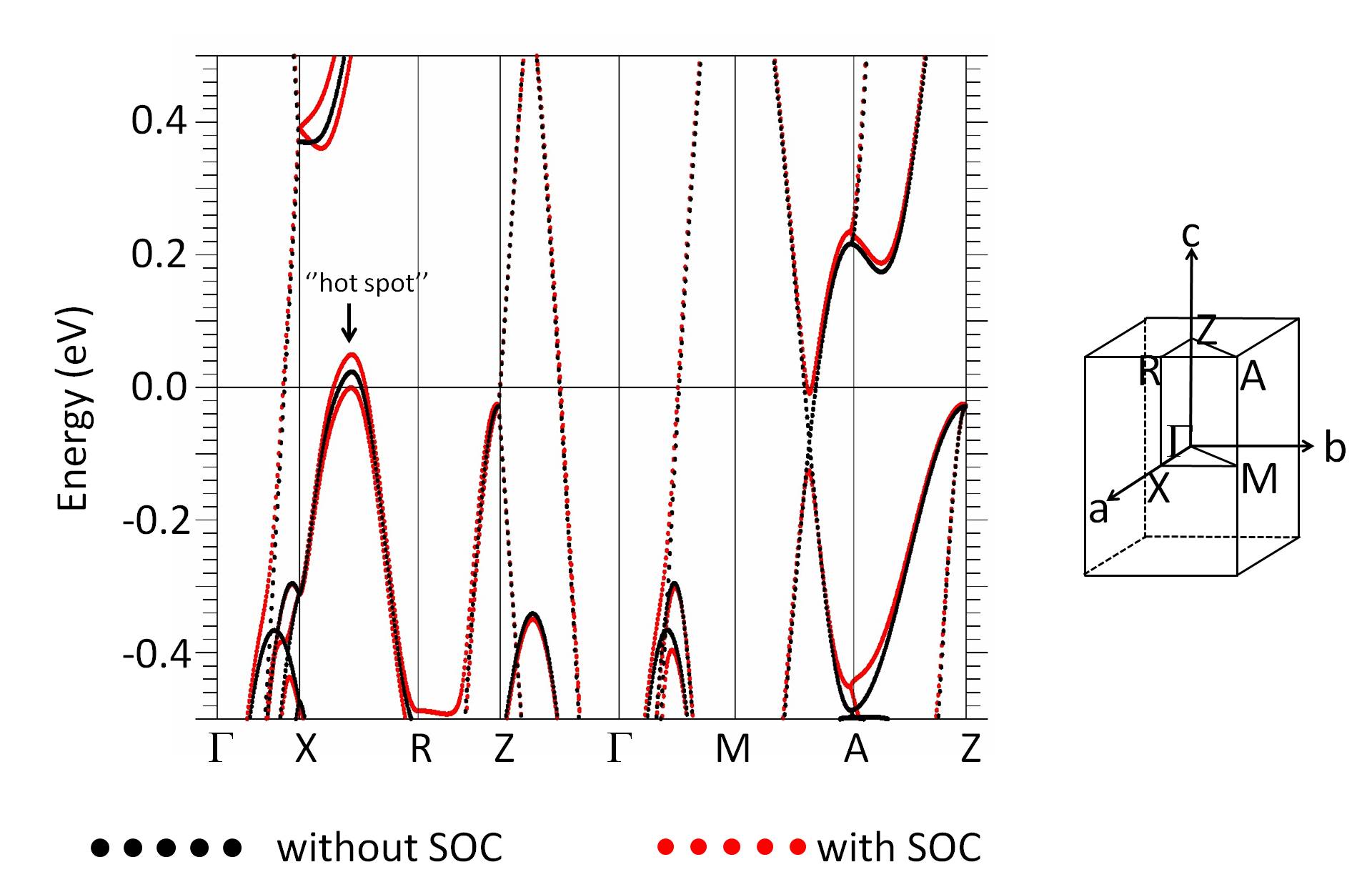}
\caption{\label{fig:figsup4}
Comparison of the non-magnetic ($U$=0) band structure without and with spin-orbit coupling (SOC).
The arrow indicates the ``hot spot'' region of the band structure, discussed in the manuscript.
}
\end{figure}

\begin{figure}[htp]
\includegraphics[width=0.6\columnwidth]{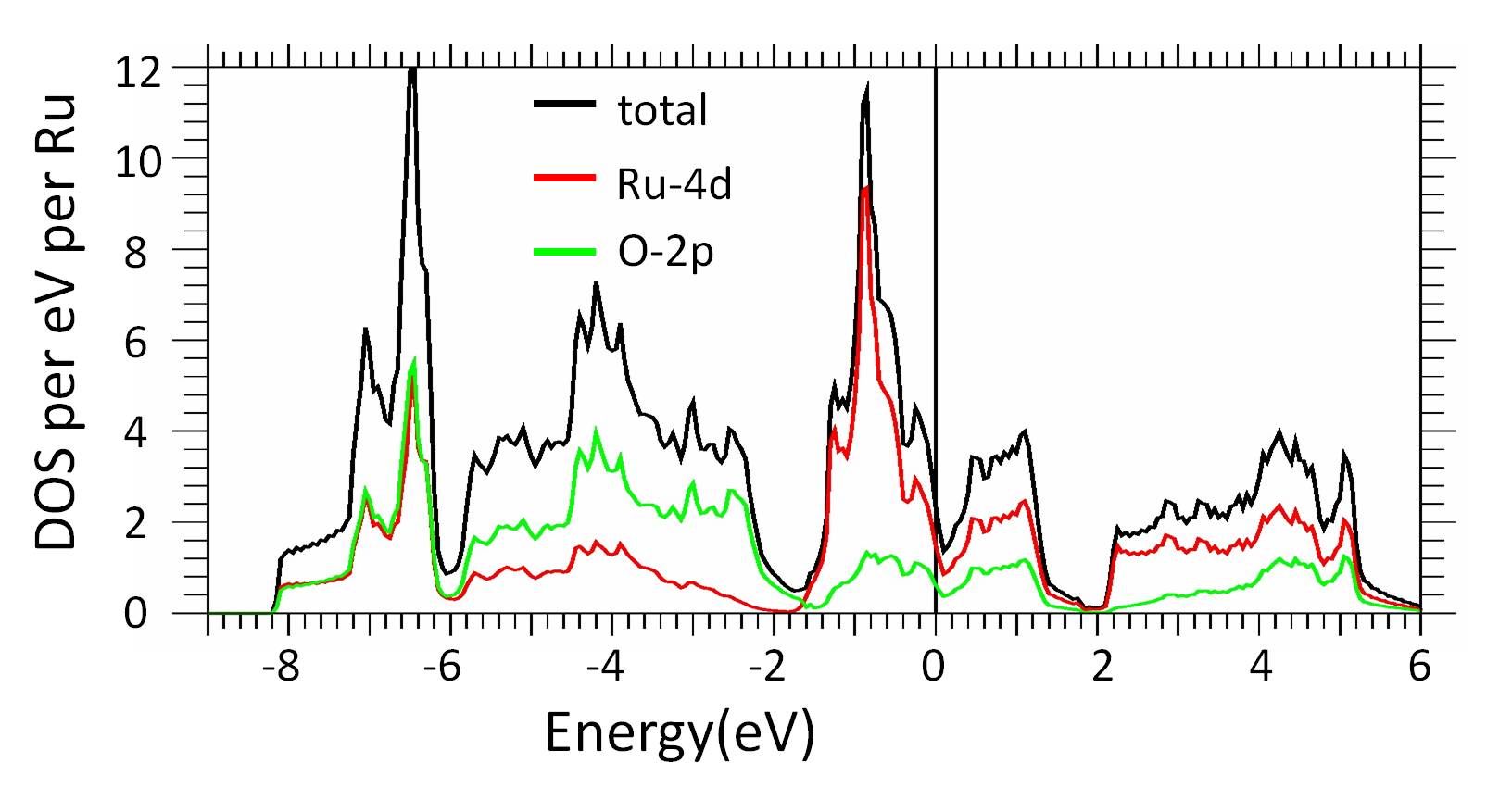}
\caption{\label{fig:figsup5}
Orbital resolved density of states from the non-magnetic PBE+$U$ simulation ($U$=2 eV).
}
\end{figure}

The orbital resolved density of states obtained from the non-magnetic PBE+$U$ simulation ($U$=2 eV) presented in Fig. \ref{fig:figsup5} shows the strong covalency between the Ru $4d$ and O $2p$ orbitals. The density of states and bandstructures in Fig. \ref{fig:figsup6} and \ref{fig:figsup7} respectively illustrate that magnetism gaps out a large portion of the Ru $t_{2g}$ band complex around the Fermi-level.

\begin{figure}[htp]
\includegraphics[width=0.6\columnwidth]{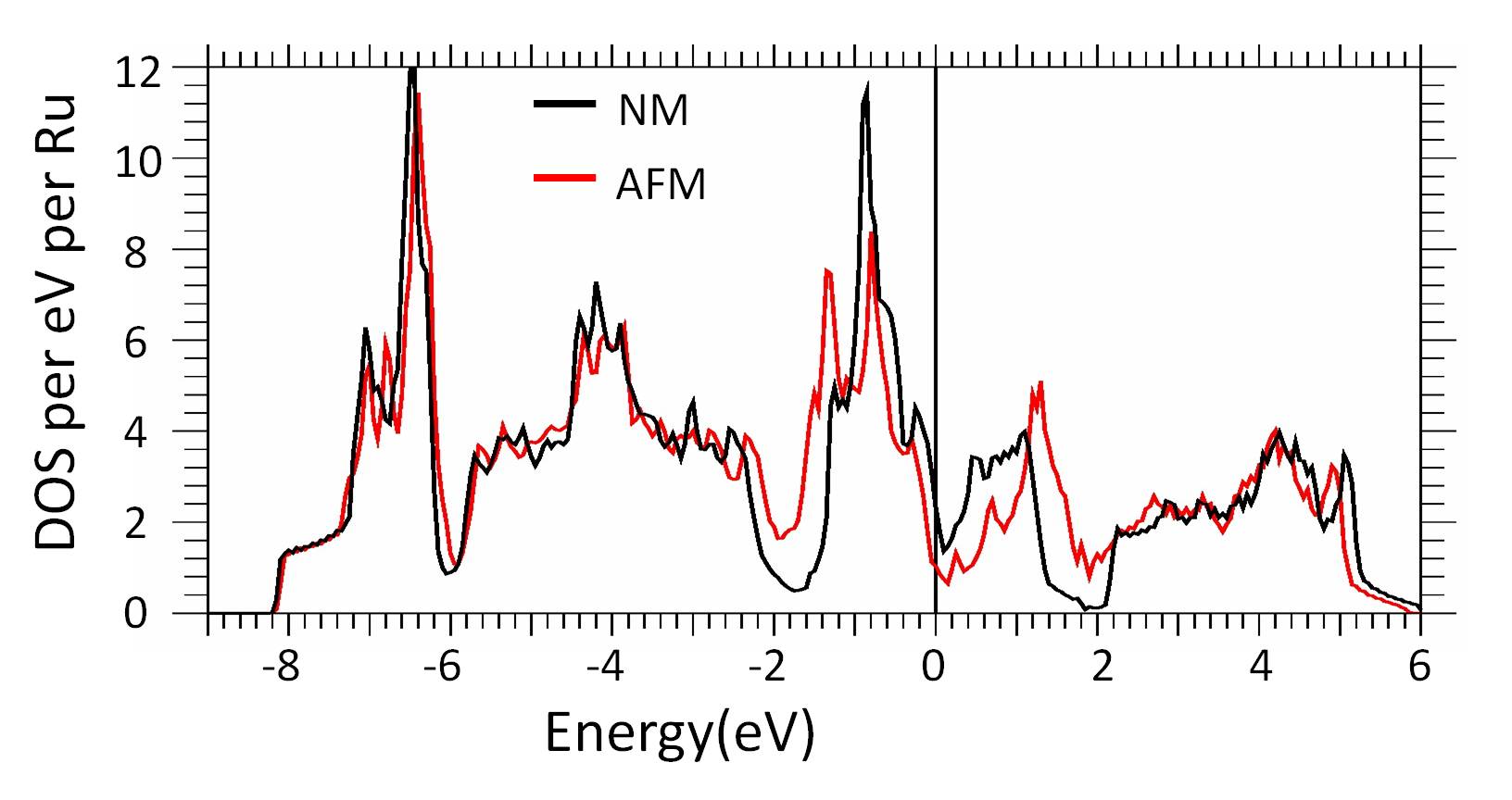}
\caption{\label{fig:figsup6}
Comparison of the non-magnetic (NM) and anti-ferromagnetic (AFM) density of states from the PBE+$U$ simulation ($U$=2 eV).
}
\end{figure}

\begin{figure}[htp]
\includegraphics[width=0.6\columnwidth]{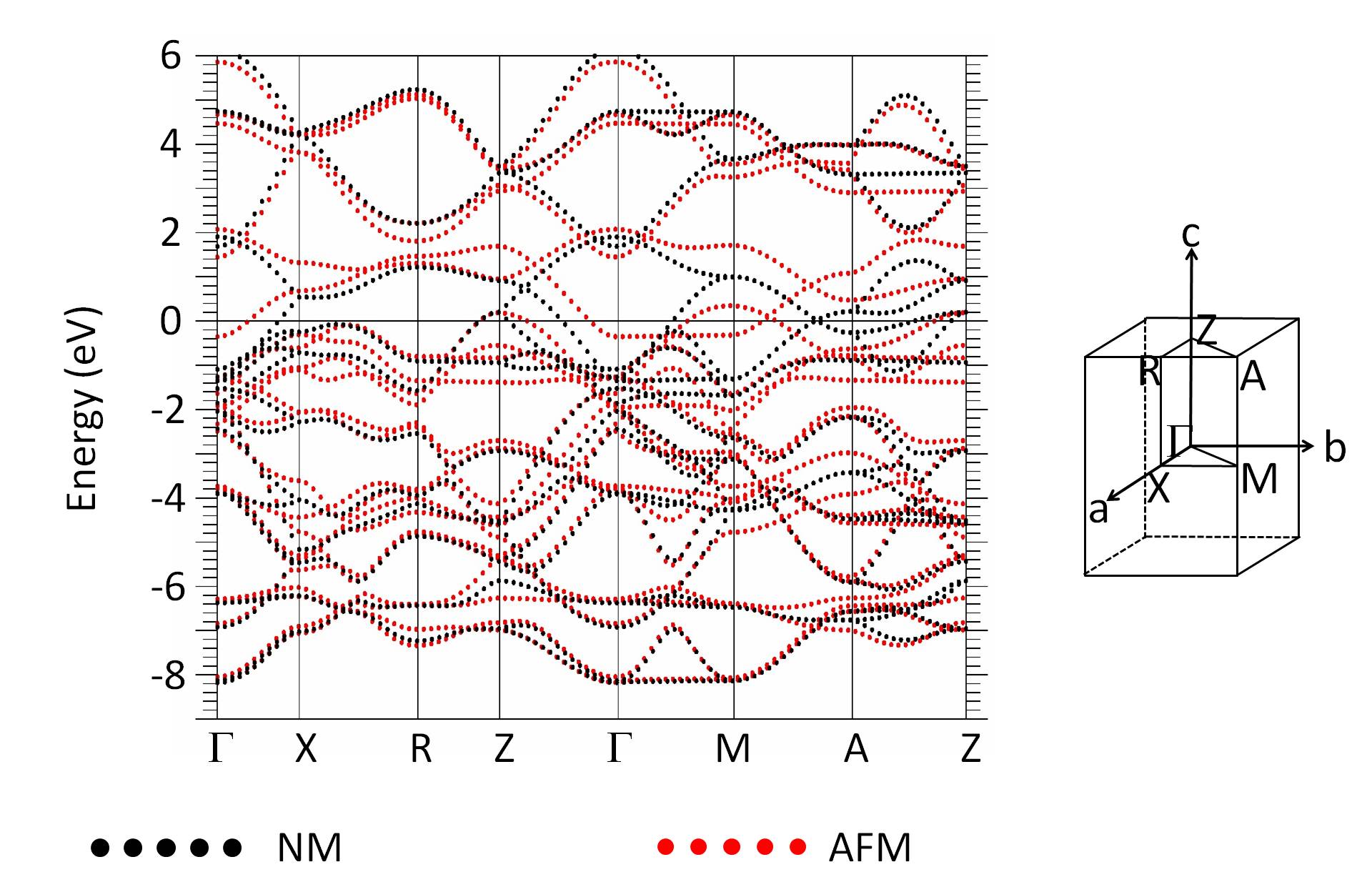}
\caption{\label{fig:figsup7}
Comparison of the non-magnetic (NM) and anti-ferromagnetic (AFM) band structures from the PBE+$U$ simulation ($U$=2 eV).
}
\end{figure}
 \clearpage

\begin{figure}[htp]
\includegraphics[width=0.6\columnwidth]{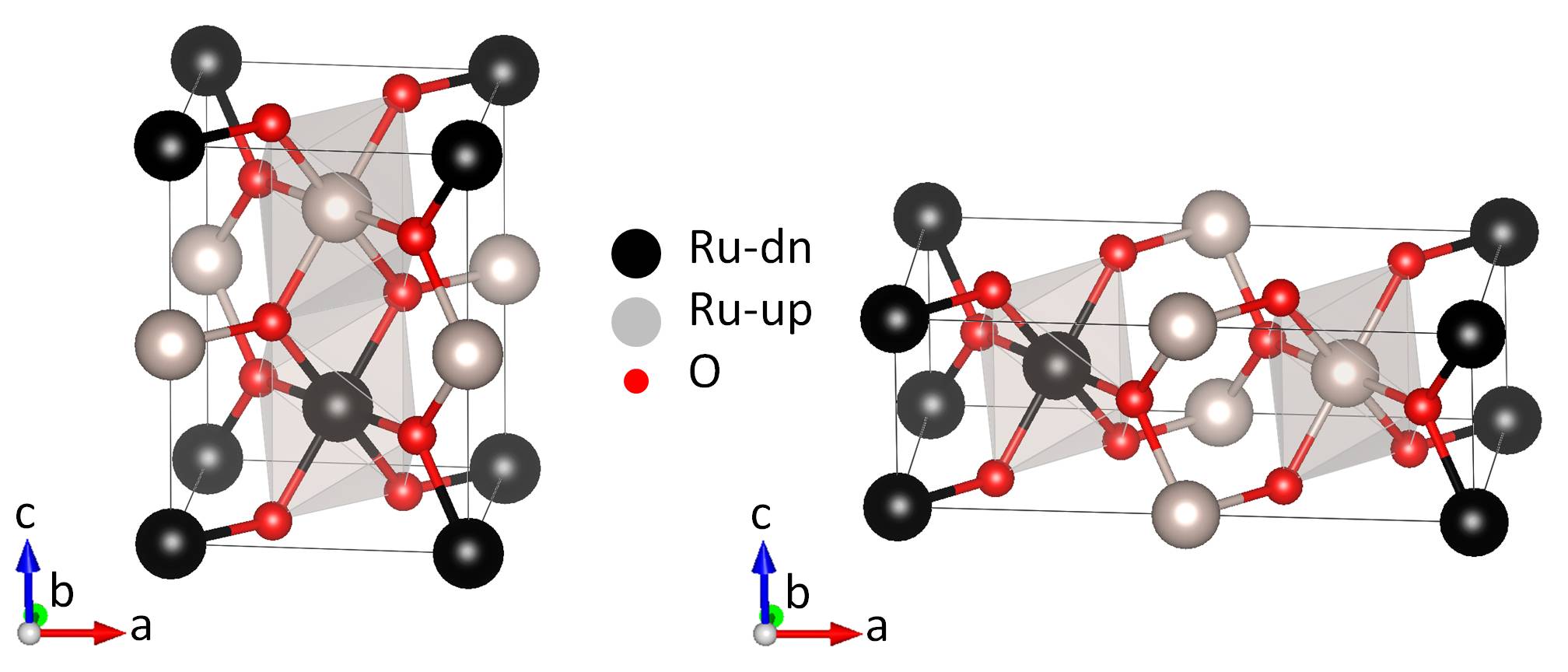}
\caption{\label{fig:figsup8}
AFM unit cell doubled in the out-of-plane (left) and in-plane (right) directions.
}
\end{figure}

 Within PBE+U with U=2eV, we explored two other AFM configurations depicted in Fig. \ref{fig:figsup8}. For the AFM configuration corresponding to doubling the unit cell in the out-of-plane/in-plane direction we found the total energy per Ru to be 11meV/107meV higher compared to that of the single unit cell AFM configuration discussed in the manuscript.

\end{widetext}

\end{document}